\documentstyle{mn}
\newread\epsffilein    
\newif\ifepsffileok    
\newif\ifepsfbbfound   
\newif\ifepsfverbose   
\newdimen\epsfxsize    
\newdimen\epsfysize    
\newdimen\epsftsize    
\newdimen\epsfrsize    
\newdimen\epsftmp      
\newdimen\pspoints     
\def\epsfbox#1{\global\def\epsfllx{72}\global\def\epsflly{72}%
   \global\def\epsfurx{540}\global\def\epsfury{720}%
   \def\lbracket{[}\def\testit{#1}\ifx\testit\lbracket
   \let\next=\epsfgetlitbb\else\let\next=\epsfnormal\fi\next{#1}}%
\def\epsfgetlitbb#1#2 #3 #4 #5]#6{\epsfgrab #2 #3 #4 #5 .\\%
   \epsfsetgraph{#6}}%
\def\epsfnormal#1{\epsfgetbb{#1}\epsfsetgraph{#1}}%
\def\epsfgetbb#1{%
\openin\epsffilein=#1
\ifeof\epsffilein\errmessage{I couldn't open #1, will ignore it}\else
   {\epsffileoktrue \chardef\other=12
    \def\do##1{\catcode`##1=\other}\dospecials \catcode`\ =10
    \loop
       \read\epsffilein to \epsffileline
       \ifeof\epsffilein\epsffileokfalse\else
          \expandafter\epsfaux\epsffileline:. \\%
       \fi
   \ifepsffileok\repeat
   \ifepsfbbfound\else
    \ifepsfverbose\message{No bounding box comment in #1; using defaults}\fi\fi
   }\closein\epsffilein\fi}%
\def\epsfsetgraph#1{%
   \epsfrsize=\epsfury\pspoints
   \advance\epsfrsize by-\epsflly\pspoints
   \epsftsize=\epsfurx\pspoints
   \advance\epsftsize by-\epsfllx\pspoints
   \epsfxsize\epsfsize\epsftsize\epsfrsize
   \ifnum\epsfxsize=0 \ifnum\epsfysize=0
      \epsfxsize=\epsftsize \epsfysize=\epsfrsize
     \else\epsftmp=\epsftsize \divide\epsftmp\epsfrsize
       \epsfxsize=\epsfysize \multiply\epsfxsize\epsftmp
       \multiply\epsftmp\epsfrsize \advance\epsftsize-\epsftmp
       \epsftmp=\epsfysize
       \loop \advance\epsftsize\epsftsize \divide\epsftmp 2
       \ifnum\epsftmp>0
          \ifnum\epsftsize<\epsfrsize\else
             \advance\epsftsize-\epsfrsize \advance\epsfxsize\epsftmp \fi
       \repeat
     \fi
   \else\epsftmp=\epsfrsize \divide\epsftmp\epsftsize
     \epsfysize=\epsfxsize \multiply\epsfysize\epsftmp   
     \multiply\epsftmp\epsftsize \advance\epsfrsize-\epsftmp
     \epsftmp=\epsfxsize
     \loop \advance\epsfrsize\epsfrsize \divide\epsftmp 2
     \ifnum\epsftmp>0
        \ifnum\epsfrsize<\epsftsize\else
           \advance\epsfrsize-\epsftsize \advance\epsfysize\epsftmp \fi
     \repeat     
   \fi
   \ifepsfverbose\message{#1: width=\the\epsfxsize, height=\the\epsfysize}\fi
   \epsftmp=10\epsfxsize \divide\epsftmp\pspoints
   \newcount\figskipcount
      \message{#1 \the\epsfysize  }
   \vbox to\epsfysize{\vfil\hbox to\epsfxsize{%
      \includegraphics{#1}%
      \hfil}}%
\epsfxsize=0pt\epsfysize=0pt}%

{\catcode`\%=12 \global\let\epsfpercent=
\long\def\epsfaux#1#2:#3\\{\ifx#1\epsfpercent
   \def\testit{#2}\ifx\testit\epsfbblit
      \epsfgrab #3 . . . \\%
      \epsffileokfalse
      \global\epsfbbfoundtrue
   \fi\else\ifx#1\par\else\epsffileokfalse\fi\fi}%
\def\epsfgrab #1 #2 #3 #4 #5\\{%
   \global\def\epsfllx{#1}\ifx\epsfllx\empty
      \epsfgrab #2 #3 #4 #5 .\\\else
   \global\def\epsflly{#2}%
   \global\def\epsfurx{#3}\global\def\epsfury{#4}\fi}%
\def\epsfsize#1#2{\epsfxsize}
\def\figinsert#1#2{\epsfbox{#1} \message{#2} }    
\begin{document}
\title[A deep $\em ROSAT$ survey - XIV. X-ray emission from faint galaxies]
{A deep $\em ROSAT$ survey - XIV. X-ray emission from faint galaxies}
\author[O. Almaini et al.]
{O.~Almaini,$^{1}$
T.~Shanks,$^2$
R.E.~Griffiths,$^3$
B.J.~Boyle,$^4$ 
N. Roche,$^5$\cr
I.~Georgantopoulos,$^6$ 
and G.C.~Stewart$^6$ \\
$^1$Institute of Astronomy, Madingley Road, Cambridge, CB3 OHA \\
$^2$Department of Physics, University of Durham, South Road,
Durham, DH1 3LE \\
$^3$Department of Physics, Carnegie Mellon University, 5000 Forbes
Ave., Pittsburgh, PA 15213, USA\\
$^4$Anglo-Australian Observatory, PO Box 296, Epping, NSW2121, Australia\\
$^5$Johns Hopkins University, Homewood Campus, 
Baltimore MD21218, USA. \\
$^6$Department of Physics, University of Leicester, University of 
Leicester, LE1 7RH }

\date{MNRAS In Press}
\maketitle

\begin{abstract}

We present a cross-correlation analysis to constrain the faint galaxy
contribution to the cosmic X-ray background (XRB).  Cross-correlating
faint optical galaxy catalogues with unidentified X-ray sources from 3
deep $\em ROSAT$ fields we find that $B<23$ galaxies account for
$20\pm7 \% $ of all X-ray sources to a flux limit of
$S(0.5-2.0$\,keV$)=4\times10^{-15}$erg$\,$s$^{-1}$cm$^{-2}$. To probe
deeper, galaxies are then cross-correlated with the remaining
unresolved X-ray images. A highly significant signal is obtained on
each field. Allowing for the effect of the $\em ROSAT$ point-spread
function, and deconvolving the effect of galaxy clustering, we find
that faint $B<23$ galaxies directly account for $23\pm3 \% $ of the
unresolved XRB at 1keV.  Using the optical magnitude of faint galaxies
as probes of their redshift distribution, we find evidence for strong
evolution in their X-ray luminosity, parameterised with the form $L_x
\propto (1+z)^{3.2 \pm 1.0}$. Extrapolation to $z=2$ will account for
$40\pm10 \% $ of the total XRB at 1keV. The nature of the emitting
mechanism in these galaxies remains unclear, but we argue that
obscured and/or low luminosity AGN provide the most plausible
explanation.

\end{abstract}


\section{Introduction}

The nature and origin of the cosmic X-ray background remains a major
unsolved problem.  The most significant progress has been made in the
soft X-ray band below 3keV since the launch of high resolution imaging
satellites such as $\it Einstein$ and $\it ROSAT$.  By resolving as
many sources as possible in the deepest $\it ROSAT$ exposures
(Hasinger et al. 1993) up to 70$\%$ of the $0.5-2$\,keV XRB has now
been resolved into discrete sources.

Using a survey of 7 deep (21-57ks) $\it ROSAT$ fields we have detected
over 400 X-ray sources above a 4$\sigma$ threshold to an approximate
flux limit of
$S(0.5-2.0$\,keV$)\sim4\times10^{-15}$erg$\,$s$^{-1}$cm$^{-2}$. By
optical spectroscopy we then identify the optical counterparts to
these X-ray sources. This technique has shown that broad-line QSOs
directly account for at least 30$\%$ of the total $0.5-2$\,keV XRB
flux (Shanks et al. 1991). As a larger sample of QSOs became
established, detailed studies of the QSO X-ray luminosity function
(Boyle et al. 1994) and the source number count distribution
(Georgantopoulos et al. 1996) have shown that QSOs are unlikely to
contribute more than 50$\%$ of the XRB at 1keV. The existence of a new
X-ray emitting population was postulated.

A further problem in explaining the XRB with QSOs is the shape of
their X-ray spectra. QSOs show relatively steep X-ray spectra with
indices of $\Gamma=2.2\pm0.1$ while the $1-10$\,keV XRB has a
significantly flatter spectrum with $\Gamma=1.4$ (Gendreau et al
1995).  Deep $\it ROSAT$ surveys have begun to resolve a population of
harder X-ray sources at the faintest flux limits (Hasinger et
al. 1993, Vikhlinin et al. 1995, Almaini et al. 1996). There have been
suggestions that this hardening may be due to a change in the
intrinsic properties of QSOs at high redshift or the effect of
intervening photoelectric absorption (Morisawa \& Takahara 1990,
Vikhlinin et al. 1995) but recent work by Almaini et al. (1996) found
no significant change in the $0.5-2\,$keV X-ray spectra of QSOs with
flux or redshift.  The spectral hardening is due to another, largely
unidentified X-ray population. Many of these unidentified X-ray
sources appear to be associated with faint galaxies, with implied
X-ray luminosities typically 100 times higher than normal field
galaxies. In particular, a number of individually identified
narrow-emission line galaxies show spectra significantly harder than
QSOs, more consistent with the spectrum of the XRB (Almaini et
al. 1996).  Similar results were obtained by Carballo et al. (1995)
and Romero-Colmenero (1996).

The nature of the X-ray emitting mechanism in these galaxies remains
unclear. This will be discussed further in Section 4, but it should be
emphasized that many of these objects could contain low luminosity or
obscured AGN (see eg. Comastri et al. 1995). In this sense the entire
XRB could still be due to `AGN' rather than two distinct classes of
X-ray source. Whatever the source of X-ray emission, it has been known
for some time that objects classified as galaxies could make a
significant contribution to the XRB.  Using $\em Einstein$
observations of local galaxies, Giacconi \& Zamorani (1987) extrapolated
the observed X-ray to optical flux ratios to the faintest ($B<27.5$)
optical counts of Tyson et al. (1988) and found that normal galaxies
could produce at least 13$\%$ of the XRB at 2keV.  Griffiths and
Padovani (1990) used the observed evolution in the 60$\mu$m luminosity
of IRAS galaxies and the local X-ray to infrared ratios to estimate
that IRAS and starburst galaxies seen to high redshift could produce
10-30$\%$ of the 0.5-3keV XRB.

Other studies have probed the galaxy contribution using the two-point
cross-correlation between the hard (2-10\,keV) XRB and various optical
and infra-red galaxy catalogues (Lahav et al. 1993, Miyaji et
al. 1994, Carrera et al. 1995, Refregier et al. 1997). These galaxy
catalogues were generally fairly shallow ($z<0.1$) but by correcting
for the effects of clustering and extrapolating the measured volume
emissivity to z$\sim$5 it was shown that 10-30$\%$ of the XRB could be
explained by sources associated with galaxies. However because of the
unknown error in such an extrapolation and the uncertain evolutionary
properties of galaxies it is clearly desirable to probe more distant
objects directly. The clearest evidence that fainter galaxies could be
important contributors to the XRB came from the study of Roche et al
(1995). Using the improved sensitivity and angular resolution of the
$\em ROSAT$ satellite they were able to probe fainter limits and much
smaller angular scales ($\sim15''$).  By cross-correlating faint
galaxy catalogues with X-ray sources they obtained a $2-3 \sigma$
detection implying that $B<21$ galaxies account for $\sim 5\% $ of the
X-ray sources.  However they obtained a more significant ($\sim
5\sigma $) signal by removing the X-ray sources and cross-correlating
with 3 deep (21-49ks) images of the $\em unresolved$ XRB. The results
implied that galaxies to a limit of $B=23$ directly contribute $\simeq
17\%$ of the 1keV XRB. A simple extrapolation to $B<28$ showed that
galaxies could account for $\sim30\%$ of the XRB or possibly more with
evolution. Similar results have since been obtained by 
 Roche et al. (1996b) and Soltan et al. (1997).

In this paper we perform an independent test of these results on 2 new
deep ($\sim$50ks) $\em ROSAT$ exposures and for the first time  attempt to
measure the evolution  in the X-ray emissivity of faint galaxies with
redshift. The deepest field (GSGP4) from the Roche
et al. (1995) work is  included, but this time the effect of
clustering is analysed more rigorously. First we will perform a
cross-correlation of the unidentified X-ray sources with faint
galaxies. We will then probe deeper into the remaining unresolved
XRB. To deconvolve the effect of galaxy clustering we will make use of
the method developed by Treyer \& Lahav (1996) (hereafter TL96),
extending their formalism to take account of the $\em ROSAT$ point
spread function.  We will then consider the cross-correlation signal
as a function of magnitude to obtain an estimate of the evolution of
X-ray emissivity with redshift.  We assume the cosmology $q_o=0.5 $\,
and $ \Lambda=0 $ throughout.

\section{Cross-correlating X-ray sources with faint galaxies}

\subsection{Data and method} 

\begin {table}
\begin {center}
\begin {tabular}{|c c c c c|}
\hline
Field & RA & DEC & $N_{H}$ & Exposure  \\ 
\hline
BJS855 & ${\rm10^h46^m24^s }$ & ${\rm -00^{\circ}21'00''}$ & $ 2.9 \pm
0.4$ & 57147s \\ 
BJS864 & ${\rm13^h43^m43^s }$ & ${\rm
-00^{\circ}15'00''}$ & $ 2.6 \pm 0.3$ & 52466s \\ 
GSGP4 & ${\rm00^h57^m29^s }$ & ${\rm
-27^{\circ}38'13''}$ & $ 1.8 \pm 0.3$ & 48995s \\ 
\hline
\end{tabular}
\end{center}
\caption {Summary of deep $\em ROSAT$ fields, with coordinates in J2000 and
galactic column density in $10^{20}\,$atom$\,$cm$^{-2}$.} 
\end{table}

In this paper we use 3 deep $\em ROSAT$ PSPC exposures with optical
identifications from the X-ray source catalogue of Shanks et al.
(1997). We refer to this catalogue paper for full details of the X-ray
source detection and optical follow-up observations.  To a flux limit
of $S(0.5-2.0$\,keV$)\sim4\times10^{-15}$erg$\,$s$^{-1}$cm$^{-2}$,
approximately $50 \%$ of the X-ray sources were identified as QSOs by
optical spectroscopy (Boyle et al. 1994).  Many of the remaining X-ray
sources had no obvious QSO candidate within their error box, but
appeared to be associated with faint galaxies on deep photographic
plates (Georgantopoulos et al. 1996).  Due to the high surface density
of galaxies at faint magnitudes however ($\sim$10000 deg$^{-2}$ at
B$<$23, Metcalfe et al. 1995) and the relatively large X-ray error
circle ($\sim$25$''$ FWHM) many of these will be chance
associations. If we look deep enough we can find a galaxy counterpart
to any X-ray source. We will therefore adopt a statistical approach to
determine the fraction of galaxies among the X-ray sources.

The 3 deepest $\em ROSAT$ fields from the survey will be used in this
analysis, including the GSGP4 field analysed by Roche et al. (1995). A
summary of these exposures is given in Table 1.  The optical galaxy
catalogues were obtained from COSMOS plate scans of Anglo-Australian
Telescope photographic plates, calibrated with magnitudes and a
star-galaxy classification flag. This optical data was previously used
for the galaxy clustering and number count analyses of Stevenson et al
(1985) and Jones et al. (1991).  Further details of the plate
reduction and zero point magnitude calibration are given in Jones et
al (1991).  Astrometric solutions to convert the plate co-ordinates
into R.A. and Dec.  were already in existence from the UVX QSO survey
of Boyle et al. (1990) and Boyle, Jones \& Shanks (1991). A second
transformation was then required to convert R.A. and Dec. into the
(x,y) $\em ROSAT$ pixel coordinates. This was performed using
approximately 10-15 bright stars and UVX QSOs with definite
counterparts on each $\em ROSAT$ image. These gave 6 coefficient
astrometric transforms with rms positional errors of between 7.1 and
8.4 arcseconds, as expected given the angular resolution of the $\em
ROSAT$ PSPC. On Figure 1 we show the resulting positions of $B<23$
galaxies on one of the fields (BJS855) in X-ray coordinates. The plate
edges and larger plate holes are clearly visible, where bad data has
been flagged around bright objects or in some case where plate defects
were detected in the plate measuring process (Jones et al. 1991).

\begin{figure}
\centering
\centerline{\epsfxsize=6truecm 
\figinsert{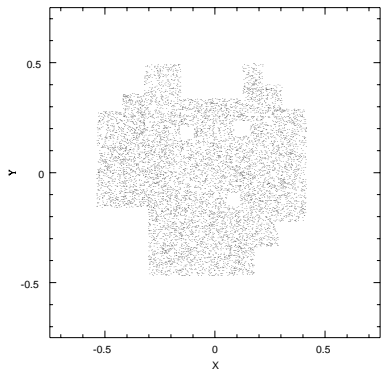}{0.0pt}}
\caption{Showing the positions of 8714 galaxies with magnitudes $B<23$ on the 
BJS855 field after transformation to X-ray coordinates (scaled to
degrees). }
\end{figure}

In calculating the cross-correlation of optical positions with X-ray
sources the number of galaxies were counted in successive annuli of
5$''$ width around each X-ray source. This distribution was then
compared with the counts obtained by placing 50,000 random points over
each of the field areas, taking care to avoid the plate edges and
``holes''. A two point cross-correlation function $C_{xo}(\theta)$ of
X-ray and optical sources can then be defined as:

\begin{equation}
C_{xo}(\theta_{i})=\frac{N_{xo}(\theta_{i})N_{r}}{N_{xr}(\theta_{i})N_{o}} -1
\end{equation}

where, $N_r$ is the total number of random points, $N_o$ the
number of optical plate sources being correlated and
{$N_{xo}(\theta_{i})$ and { $N_{xr}(\theta_{i})$ give the number of
X-ray/optical and X-ray/random pairs respectively.

\subsection{Results}

\begin{figure}
\centering
\centerline{\epsfxsize=7truecm 
\figinsert{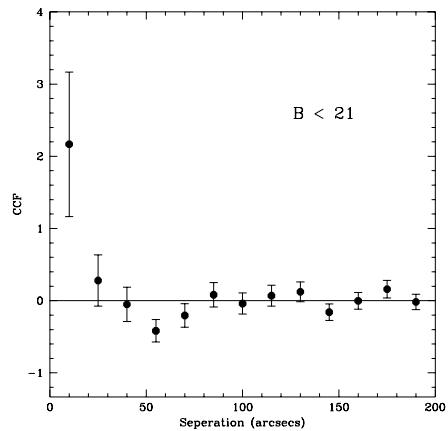}{0.0pt}}
\centerline{\epsfxsize=7truecm 
\figinsert{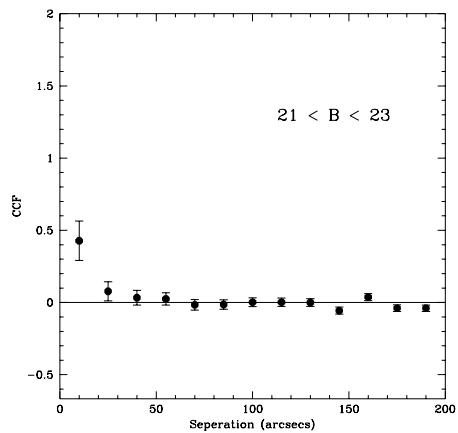}{0.0pt}}
\caption{(a) The cross-correlation function $C_{xo}(\theta)$  of unidentified
X-ray sources from 3 deep fields with $B<21$ mag galaxies.
On (b) we repeat the analysis with fainter $21\leq B \leq 23$ mag galaxies.
Note the change in scale compared to (a).
Error bars are due to $\protect\sqrt{N}$ statistics. }
\end{figure}

\begin {table*}
\begin {center}
\begin {tabular}{|c|c c c| c|}
\hline
Magnitude range & Field & Pairs $<30''$ &  Expected & Excess significance \\
\hline
                   & GSGP4 & 22            &  13.2    & 2.4$\sigma$\\
      $B<21$       & BJS855  & 19            &  12.3     & 1.9$\sigma$\\
                   & BJS864  & 11            &  5.93     & 2.1$\sigma$\\
\cline{2-5}
                   & Total & 52            &  31.5     & 3.7$\sigma$\\
\hline
                   & GSGP4 & 123            &  104.9     & 1.8$\sigma$\\
$21\leq B \leq 23$ & BJS855  & 101            &  85.0     & 1.7$\sigma$\\
                   & BJS864  & 102           &  88.1     & 1.5$\sigma$\\
\cline{2-5}
                   & Total & 326           &  278    & 2.9$\sigma$\\
\hline
\end{tabular}
\end{center}
\caption {
Showing the number of faint galaxies found within $30''$ of 
unidentified X-ray sources on 3 deep fields compared to the number
expected by chance.  The galaxies are split into those with $B<21$ and
fainter galaxies with $21\leq B \leq 23$. }
\end{table*}

Removing only the X-ray sources identified with QSOs or galactic stars
we cross-correlate the positions of the remaining 149 unidentified $4
\sigma$ sources with the objects classified as galaxies on
photographic plates. These optical galaxies were split into a bright
sample with $B<21$ and a fainter sample with $21\leq B\leq23$. The
results are shown in Table 2, where we display the number of galaxies
found within 30$''$ of an X-ray source compared to the distribution
expected by chance. A 30$''$ radius approximately corresponds to the
3$\sigma$ $\em ROSAT$ PSPC positional error.

Considering first the cross correlation with brighter $B<21$ galaxies,
on each field we find an excess compared to a random distribution.
Overall we find 52 source-galaxy pairs compared to 31.5 expected by
chance, amounting to a 3.7$\sigma$ rejection of the null hypothesis
that the two distributions are not correlated. This would indicate an
excess of $\simeq20.5\pm 7.2$ $B<21$ galaxies around X-ray sources. On
Figure 2(a) we display the source-galaxy cross-correlation in 15$''$
radial bins.

To probe deeper, we repeat the cross-correlation with the $\sim$5000
fainter galaxies in the magnitude range $21\leq B \leq 23$. We find
326 galaxies within 30$''$ of an X-ray source compared to 278
expected. This represents a 2.9$\sigma$ rejection of the hypothesis of
no correlation and indicates an excess of $\simeq48\pm 17$ of these
fainter galaxies around X-ray sources. The cross-correlation function
is displayed in Figure 2(b) for the 3 fields combined.

We must now make a correction for the effect of galaxy clustering.
Since galaxies are clustered on the scales probed by this analysis
this can  enhance the excess of source-galaxy pairs. We therefore
apply a simple correction based on $\omega_{gg}(\theta)$, the angular
autocorrelation function of faint galaxies (from Roche et al
1996a). Defining $\Delta N'_{gx}(\theta)$ to be the observed excess of
source-galaxy pairs measured within a  given angle $\theta$, this will
be enhanced relative to the true number of X-ray emitting galaxies
$\Delta N_{gx}(\theta)$ such that:

\begin{equation}
\Delta N_{gx}(\theta) = \frac{\Delta N'_{gx}(\theta)} {1 + G_{tot}^{-1}
\Delta N_{gg}(\theta)}
\end{equation}

where $\Delta N_{gg}(\theta)$ is the excess of galaxies within angle $\theta$
compared to a random distribution, obtained by integration of
$\omega_{gg}(\theta)$, and $G_{tot}$ is the total number of galaxies in
the entire image. For the brighter $B<21$ galaxies this gives a $25\%
$ clustering enhancement within a radius of 30$''$.  At $21<B<23$ the
effect is slightly more significant, accounting for $31\% $ of the
excess signal.

Thus overall, allowing for the effects of clustering, we have evidence
that $\simeq48\pm17$ of the 149 unidentified X-ray sources are due to
galaxies with $B<23$, strengthening the initial findings of Roche et
al (1995). Hence from a total of 236 sources on 3 deep fields we have
shown that faint ($B<23$) galaxies contribute $\sim 20 \pm 7 \%$ of
the total source counts to a flux limit of $S(0.5-2.0
$keV$)\sim4\times10^{-15}$erg$\,$s$^{-1}$cm$^{-2}$.

However when we scale this $20 \%$ galaxy contribution by the median
flux of the unidentified X-ray sources
($\sim7\times10^{-15}$erg$\,$s$^{-1}$cm$^{-2}$) we obtain a total
contribution of $1.2 \pm 0.4\times 10^{-9} $ erg s$^{-1}$ cm $^{-2}$
sr$^{-1}$ which accounts for only $\sim 4\%$ of the total XRB. We will
therefore attempt to probe much deeper in Section 3 by
cross-correlating with the remaining $\em unresolved$ XRB.

\section{Cross-correlating the unresolved X-ray background with faint galaxies}

\subsection{The data}

In this section we  extend the cross-correlation technique
to probe the remaining unresolved component of the cosmic X-ray
background and thus investigate the origin of X-ray emission beyond
the limit of significant source detection.  By performing the analysis
with different magnitude slices of the galaxy sample we may also be
able to deduce the redshift evolution of X-ray emissivity from faint
galaxies.

The X-ray images were obtained from deep ($\sim 50$ks) $\em ROSAT$
observations reduced using the STARLINK $\em Asterix$ X-ray data
reducing package.  Since the X-ray background below $0.5$\,keV is
dominated by galactic emission and solar scattered X-rays (Snowden \&
Freyberg 1993) the images used in the cross-correlation are extracted
from only the $0.5-2$\,keV data. Data from periods of high particle
background were also removed from the analysis, excluding
approximately $10\%$ of the photons when the Master Veto Rate was
above 170 counts s$^{-1}$ (Plucinsky et al. 1993). Additional data was
available for the BJS855 and BJS864 fields from serendipitous
pointings obtained from the $\em ROSAT$ data archive, offset 6 and 7
arcminutes respectively from the original field centres and providing
an additional $\sim 20$ks of data. This data was processed separately
before a mosaic of the final images was produced. The 4$\sigma$
sources were then removed from the data using a range of exclusion
radii suitable for excluding 98$\%$ of the photons, depending on the
off-axis angle of the source. This radius varied from 30.6$''$ for a
source on axis to 75.6$''$ for a source at a radius of 20 arcminutes.
Inspection of the resulting images revealed that only one source
showed evidence for being extended and could not be successfully
removed in this way.  This source, GSGP4X:032, was identified as a
distant ($z=0.56$) galaxy cluster. A 70$\%$ larger radius was required
to successfully remove this source. Finally, only the central 16
arcminute regions were used in the cross-correlation analysis since
the sensitivity of the PSPC drops off rapidly beyond 20 arcminutes
(Hasinger et al. 1992).  For the BJS855 and BJS864 fields, where deeper
exposures have been obtained by adding additional archive data, only
regions of the final image lying within 16 arcminutes of $\em both$
field centres are used.

\subsection{Cross-correlation results}

A two point cross-correlation between the unresolved X-ray images and
the faint galaxy catalogues was obtained by counting the number of
X-ray photons in successive annuli around each galaxy and comparing
this with the mean pixel intensity. By repeating for every galaxy in
the field (avoiding holed regions), the cross-correlation is defined
by:

\begin{equation}
W_{xg}(\theta_{i})=\frac{\sum N_{x}(\theta_{i})}
{\sum N_{p}(\theta_{i})\left\langle{N_{X}}\right\rangle} -1
\end{equation}

\noindent 
where $N_{x}(\theta_{i})$ and $N_{p}(\theta_{i})$ are the total number
of X-ray photons and pixels respectively within annulus $\em i$ of
each galaxy, $\left\langle{N_{X}}\right\rangle$ is the mean number of
photons per pixel and the summations occur over all the galaxies in
the field.  Error estimates were obtained by splitting each field into
4 quadrants and obtaining the cross-correlation for each separately.
The error on $W_{xg}$ was then estimated from the error on the mean
from these quadrants.

\begin{figure}
\centering
\centerline{\epsfxsize=6truecm 
\figinsert{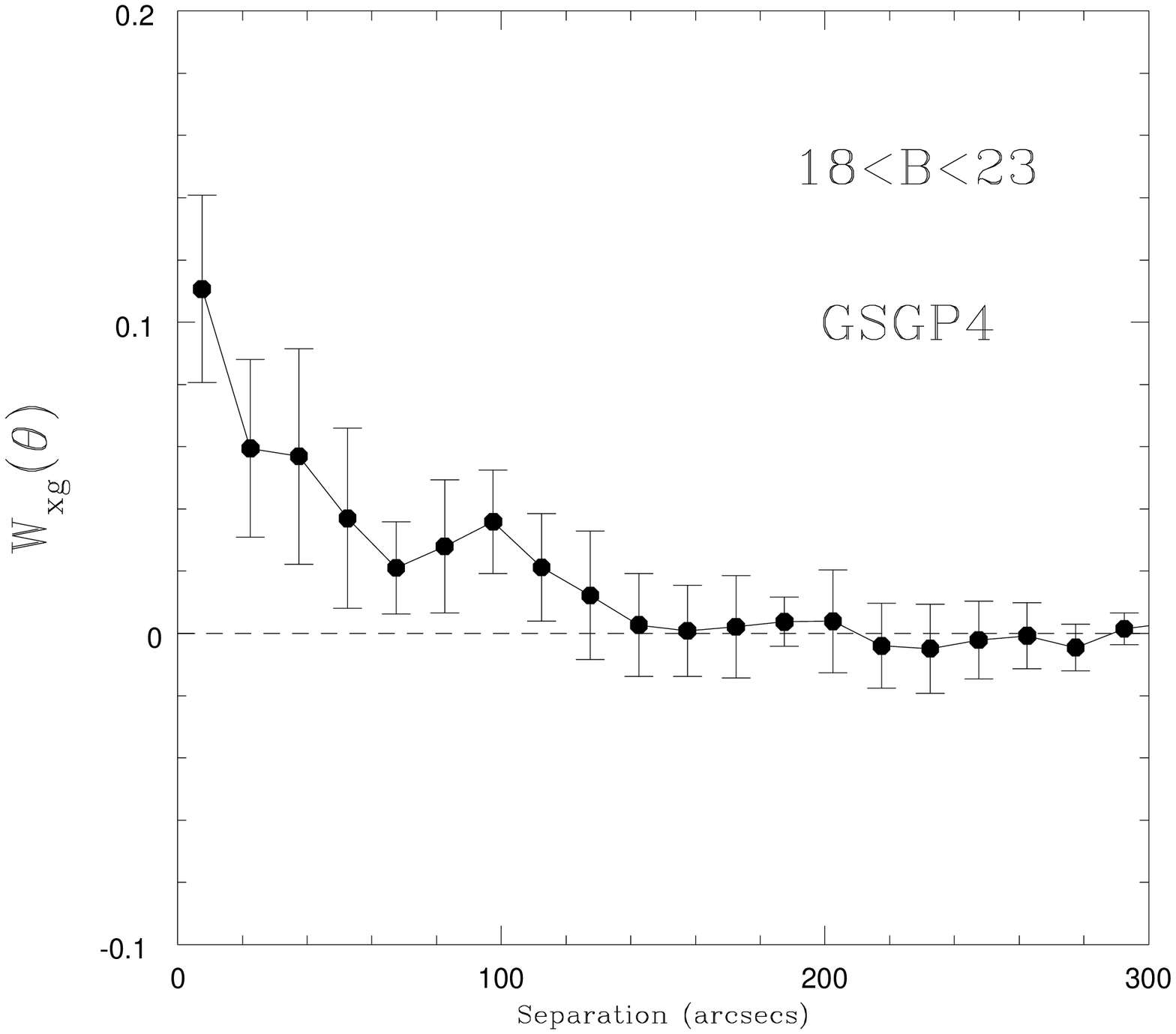}{0.0pt}}
\centerline{\epsfxsize=6truecm 
\figinsert{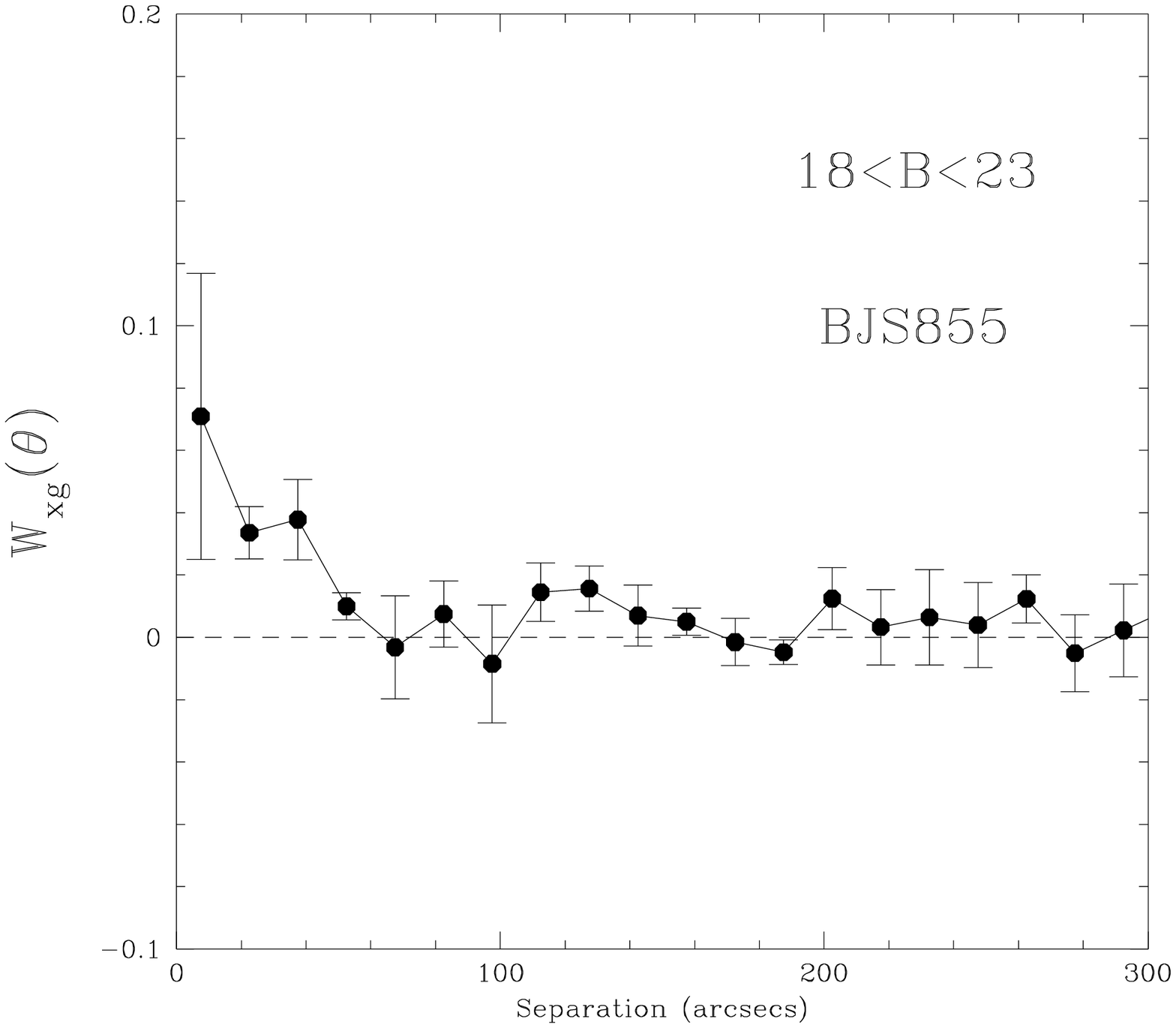}{0.0pt}}
\centerline{\epsfxsize=6truecm 
\figinsert{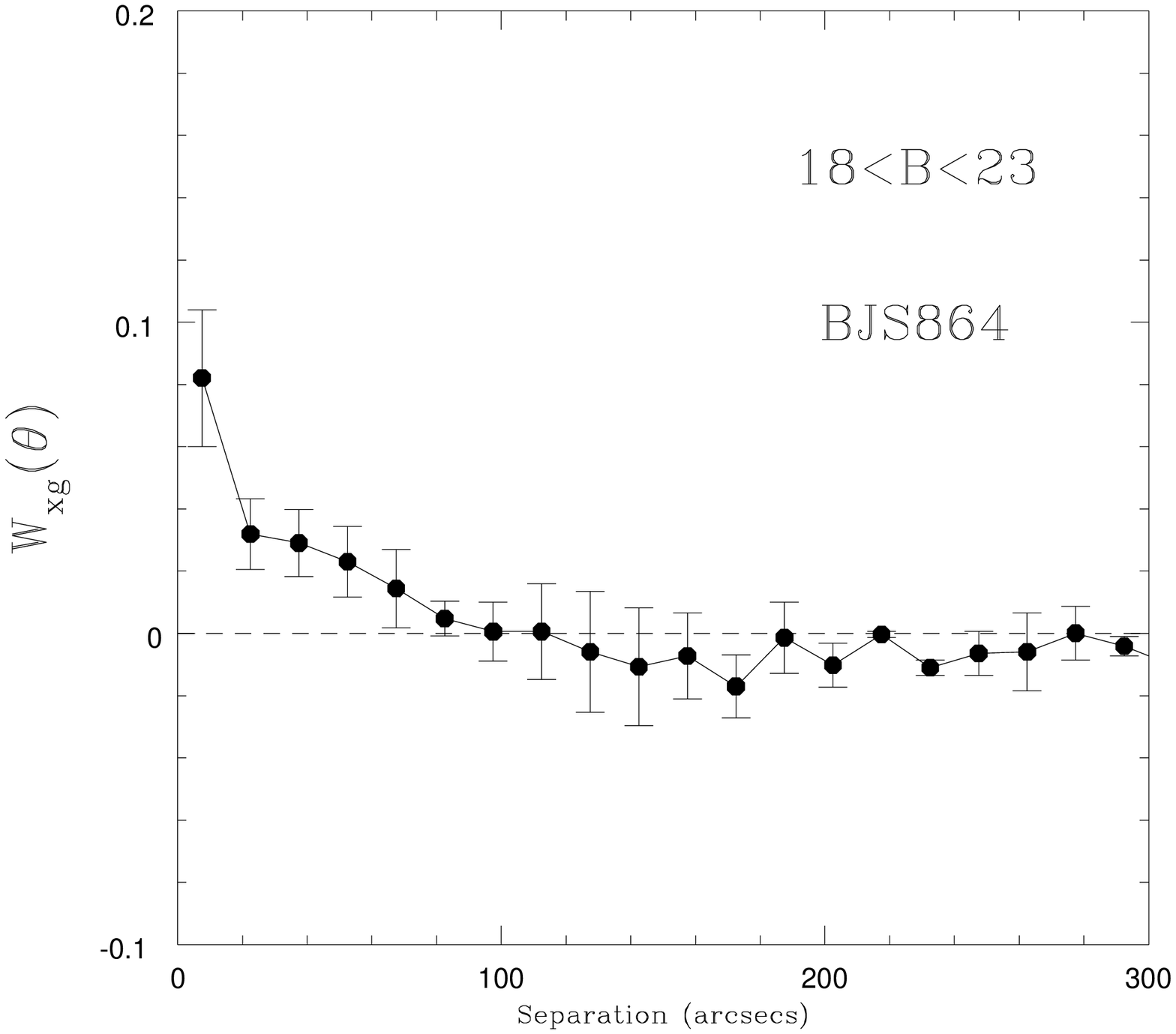}{0.0pt}}
\caption{The  cross-correlation function $W_{xg}(\theta)$  
of the unresolved $0.5-2$keV X-ray background with faint  $18\leq B \leq 23$ 
galaxies on 3 deep fields. }
\end{figure}

The resulting two point cross-correlations between faint $18<B<23$
galaxies and the unresolved XRB are shown in Figure 3 in 15$''$
bins. It is clear that a significant positive signal has been detected
on each field. The errors on the first two bins alone ($<30''$)
suggest individual detections $>3\sigma$ significance. The independent
analysis of the GSGP4 field is seen to be in excellent agreement with
the results of Roche et al. (1995). Thus we find further strong
evidence that faint galaxies are not only significant contributors to
the resolved X-ray source counts but also to the fainter, as yet
unresolved component of the X-ray background. In Table 3 we show the
number of galaxy-photon pairs within 30$''$ of the galaxies compared
to the counts expected from a random distribution. The excess X-ray
photons close to galaxies suggest $ > 3.5\sigma$ detections on
each field assuming Poisson statistics. We have clearly confirmed the
initial findings of Roche et al. (1995).

In Figure 4 we show $W_{xg}$ for the 3 deep fields combined, where
we have summed galaxy-photon pairs across all fields and calculated
the expected counts by combining the 3 separate $\sum
N_{p}(\theta_{i})\left\langle{N_{X}}\right\rangle$ terms.  Errors have
been estimated in the same manner as before  by calculating 4
cross-correlations from the quadrants of  each field  and
using the error on the mean values of $W_{xg}$.

\begin {table}
\begin {center}
\begin {tabular}{|c c c| c|}
\hline
Field     &  Photons $<30''$   & Expected  &  Excess significance\\
\hline
GSGP4  &  8575     &     8154       &   4.7      \\
BJS855 &   5763     &   5490       &    3.7  \\
BJS864 &  7771  & 7414 &  4.1 \\
\hline
 Total   & 22109    &  21058  &  7.2   \\
\hline
\end{tabular}
\end{center}
\caption {
Results of the cross-correlation of $18\leq B \leq 23$ catalogue galaxies
with the unresolved $0.5-2.0$\,keV X-ray 
background on 3 deep fields,
comparing the  number of $0.5-2.0$\,keV X-ray photons found within 
30$''$ with a random distribution. }
\end{table}

\subsection{The effect of galaxy clustering}

We have established a highly significant cross-correlation between the
unresolved component of the cosmic XRB and faint galaxies.  Galaxies
do not randomly sample the sky however and are well known to show
evidence of clustering and structure on the scales probed by our
analysis. It is therefore probable that a non-negligible fraction of
the enhanced signal in the cross-correlation is due to the clustering
of galaxies with each other. In effect, the galaxies will correlate
with the emission of their neighbours as well. One approach (see Roche
et al. 1995) is to apply an approximate correction using the $\em
angular$ clustering of the observed optical galaxies. However this
will not account for the X-ray emission arising from fainter unseen
objects ($B>23$) which may be clustered with the galaxy catalogue.
This effect is discussed in detail in the work of TL96, where a
prescription for modelling these populations is presented. We will now
apply this formalism to our data, taking care to allow for the effect
of the $\em ROSAT$ PSPC point-spread-function (PSF). This effect was
neglected by TL96. A detailed description of the effect of the PSF can
be found in Refregier et al. (1997).

\begin{figure}
\centering
\centerline{\epsfxsize=10truecm 
\figinsert{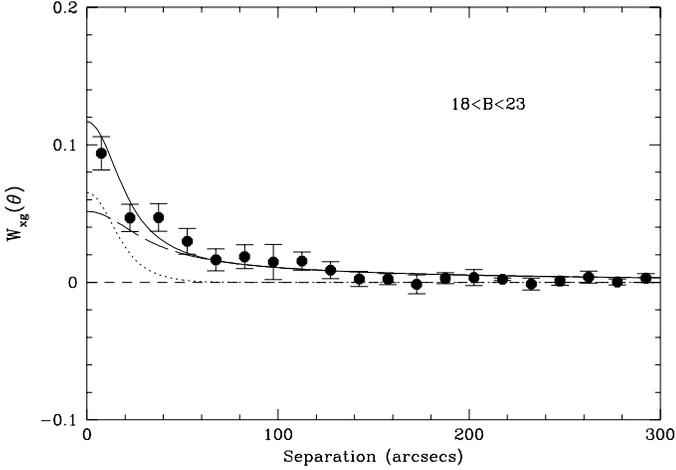}{0.0pt}}
\caption{The total cross-correlation function $W_{xg}(\theta)$ of the
unresolved $0.5-2.0$\,keV X-ray background on 3 deep $\em ROSAT$
fields with $18\leq B \leq 23$ galaxies.  Also shown is the best
fitting model of the form given in Equation 14, formed as the sum of
the Poisson term (dotted line) and a clustering term (dashed line).
}
\end{figure}

\subsection{Correlation functions and the effect of the PSF}

Under the assumption that the XRB arises from discrete, point-like
X-ray sources, it has been shown that two terms will contribute to the
cross-correlation of faint galaxies with the X-ray background (Lahav
1992, Treyer \& Lahav 1996, Refregier et al. 1997).

\begin{equation}
\eta_0 = \eta_p  + \eta_c .
\end{equation}

For the ideal case of a delta-function PSF with cell sizes $\omega
\rightarrow 0$, this function is related to the normalized form of
$W_{xg}$  (Equation 3) by:

\begin{equation}
W_{xg}(\theta)={\eta_0 \over \omega^2\bar{I}\bar{N}}
\end{equation}

In this expression $\bar{I}$ is the mean intensity of the unresolved
XRB and $\bar{N}$ gives the number of catalogue galaxies per
steradian.  The $\eta_p$ term comes from the direct contribution from
the catalogue galaxies themselves. This is known as the $\em Poisson$
term, which in the idealized case only exists at zero lag
($\theta=0$). For cells of solid angle $\omega$ this term is directly
related to the X-ray intensity $\Delta \bar I_{g}$ contributed by the
galaxies such that:

\begin{equation}
\eta_p = \omega \Delta  \bar I_{g} =
\omega \int_z {\rho_g(z)\over 4\pi r_l^2} {\rm d}v(z) .
\end{equation}

where we define $\rho_g(z)$ as the observed volume emissivity of the
galaxy population, ${\rm d}v(z)$ is the volume element per unit angle
and $r_l$ is the luminosity distance.

The second term $\eta_c$ arises from the clustering of the X-ray
sources with the galaxy population and is known as the $\em
clustering$ term. Treyer \& Lahav (1996) have calculated the
theoretical angular cross-correlation of a galaxy population with the
X-ray background. They find that the clustering term $\eta_c$ obeys a
simple power law of the form:

\begin{equation}
\eta_c(\theta)=A_{xg}\theta ^{1-\gamma}
\end{equation}

where the amplitude $A_{xg}$ can in principle be used to evaluate the
X-ray volume emissivity due to galaxies given a specific model of
their redshift distribution, clustering and luminosity evolution (see
Section 3.5). TL96 applied this simple form to the measurements of
Roche et al. (1995) to obtain an estimate of the galaxy contribution to
the XRB.  As we will demonstrate below, fitting this functional form
directly to measurements of $W_{xg}(\theta)$ isn't strictly valid due
to the effect of $\em ROSAT$ PSF. This has the effect of smearing out
the Poisson term to make a significant contribution at non-zero
lag. We will discuss how to deconvolve the Poisson and clustering
terms in Section 3.6.  First we will use the formalism of TL96 and a
specific model for the galaxy population to relate the amplitude of
the clustering term $A_{xg}$ to the X-ray volume emissivity.

\subsection{Modelling the galaxy population}

We assume that any diffuse component of the $0.5-2.0$\,keV XRB
is negligible in comparison with the source component and model the
observed volume emissivity of the source population using an
evolutionary parameter $q$ such that:

\begin{equation}
\rho_{\small XRB}(z)=\rho_s(z)=\rho_0(1+z)^q 
\end{equation}

Thus the intensity of the XRB per unit solid angle is given by:

\begin{equation}
\bar I = \omega \int_z {\rho_s (z) \over 4\pi r_l^2(z)}{\rm d}v(z)
\end{equation}

We will further assume  the spatial cross-correlation of galaxies
with the X-ray sources $\xi(r,z)$ to be the same as the
auto-correlation of faint galaxies with themselves.

Next we must characterise the properties of the galaxy population at a
given redshift.  Deep spectroscopic surveys (eg. Glazebrook et al
1995) have measured the galaxy $N(m,z)$ to optical magnitudes of
$B\simeq 24$. We hope to model even deeper than this however.  Fainter
galaxies lack redshifts, so we will use the latest evolution model of
Roche et al. (1996a). This model is consistent with current
spectroscopic results (eg. Cowie et al. 1996) and reproduces the faint
number count distribution. In general this model predicts a higher
redshift distribution for faint blue galaxies than the parametric form
of Efstathiou (1995) used in TL96.

Next we must model the clustering properties of the galaxy population
and its evolution with redshift. We will adopt the standard form
(Peebles 1980):

\begin{equation}
\xi(r,z) = (1+z)^{-(3+\epsilon)} \left(r \over r_o \right)^{-\gamma} 
\end{equation}

where the proper coordinate r is the spatial separation between the
sources and $\epsilon$ models the clustering evolution.  We will adopt
a stable clustering model ($\epsilon=0$) and a correlation radius of
$r_0=4.4~h^{-1}$Mpc (Loveday et al. 1995). Using these parameters and
the galaxy evolution model described above, Roche et al. (1996a) were
able to fit the observed galaxy angular clustering to a magnitude of
$B=27$.

To interpret our observations we will use the framework developed by
TL96, who have calculated the theoretical angular cross-correlation of
a galaxy population with the X-ray background. The detailed
calculations are somewhat cumbersome and will not be repeated
here. However they find that the clustering term $\eta_c(\theta)$
obeys a simple power law (Equation 7) where the amplitude is given by:

\begin{equation}
 A_{xg} = { \rho_o r_o^\gamma H_\gamma f(\aleph) \over
 4\pi }
\end{equation}

The function $H_\gamma$ is given by:

\begin{equation}
 H_\gamma=\int_{-\infty}^{+\infty}
dx(1+x^2)^{-\gamma/2}
\end{equation}

Assuming the standard $\gamma=1.8$ leads to  $H(\gamma)=3.68$.
Defining a global evolution parameter $\aleph = \gamma-\epsilon+q-5$,
the function $f(\aleph)$ takes the form:

\begin{equation}
f(\aleph)=\int_z {\rm d} z (1+z)^{\aleph} r_c^{1-\gamma} \int_{B_min}^{B_max} 
N(m,z)\rm{d} m.
\end{equation}

where $B_{min}$ and $B_{max}$ represent the magnitude range
of the galaxy catalogue. 

Thus we may in principle re-arrange Equation 11 to obtain a value for
the local X-ray volume emissivity $\rho_o$ using our chosen
description of the galaxy population and our observed X-ray parameters
$\bar{I}$ and $A_{xg}$. As discussed above however, measuring $A_{xg}$
is non-trivial since we must deconvolve the effect of the Poisson
term. This will be discussed in the next section.

In order to estimate the evolutionary parameter $\aleph$ we need a
value for the parameter $q$ which describes the evolution in the $\em
observed$ X-ray emissivity (Equation 8).  We will attempt to
measure this evolution directly in Section 3.7.

\subsection{Deconvolving the Poisson and clustering terms in $W_{xg}(\theta)$}

We define the PSF of the $\em ROSAT$ PSPC to be the function
$\psi(\theta)$ normalised to unity over the whole sky. We adopt the
parametric form given by Hasinger et al. (1995). Our images are
extracted from the central 16 arcminute region of the PSPC and we will
use the mean weighted PSF over this radius, assuming a mean photon
energy of 1keV.

The effect of the $\em ROSAT$ PSF is to smear out the X-ray emission
over neighbouring pixels, and hence the Poisson and clustering terms
in the cross-correlation, derived assuming a delta function PSF, will
be smeared out compared to Equation 5 to produce the measured
$W_{xg}(\theta)$ as follows:

\begin{eqnarray}
W_{xg}(\theta_{12}) & = & { 1 \over \bar{I}\bar{N}2\pi\omega^2} \int_0^{2\pi}
\rm{d}\phi_{12} 
\int_{C_1}
\rm{d}\Omega_{1} 
\int_{C_2}
\rm{d}\Omega_{2} 
\nonumber
\\
& & \times \int_S \rm{d}\Omega_S 
\,\psi(\theta_{2s})\,\eta(\theta_{1s})
\end{eqnarray}

The $\rm{d}\Omega_{1}$ and $\rm{d}\Omega_{2}$ integrals are evaluated
over cells $C_1$ and $C_2$ separated by $\theta_{12}$.  A
rigorous derivation of such equations can be found in Refregier et al. (1997). The final integral is evaluated  over the whole sky,
and is effectively a 2-dimensional convolution of the point spread
function with the idealized correlation function $\eta(\theta)$.

Although the two terms $\eta_c$ and $\eta_p$ both contribute to the
observed $W_{xg}(\theta)$ they are not independent.  By Equation 6 the
Poisson term $\eta_p$ can be evaluated from the X-ray volume
emissivity, which in turn is related to the clustering term $\eta_c$
via Equations 8 and 11. Therefore $ \eta(\theta)$ in the above integral
may be replaced by:

\begin{equation}
\eta(\theta) = A_{xg}\left[ \theta^{1-\gamma} + {4\pi \over
  r_o^{\gamma}H_{\gamma}f(\aleph)} \int_z {(1+z)^q \over 4\pi r_l^2}
  {\rm d}v \right]
\end{equation}

We must therefore evaluate this functional form of $W_{xg}(\theta)$
numerically and then fit to our observations to obtain an estimate of
the amplitude $A_{xg}$.  In order to evaluate this function, however,
we must assume a value for the evolutionary parameter $q$. We will now
attempt to measure this evolution directly.

\subsection{Constraining evolutionary parameters}

\begin{figure}
\centering \centerline{\epsfxsize=8truecm \figinsert{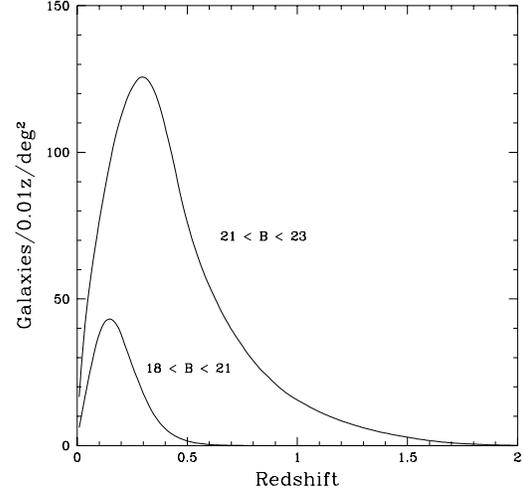}{0.0pt}}
\caption{Galaxy redshift distributions predicted using the PLE model
of Roche et al. (1996a) for $B<23$ galaxies split into a `bright' and
`faint' sample.}
\end{figure}

The largest uncertainty in previous estimates of the galaxy
contribution to the XRB has been the assumed redshift evolution
(eg. see Lahav et al. 1993, Roche et al. 1995).  Treyer \& Lahav (1996)
proceed under the assumption that the luminosity ratio $L_X/L_{opt}$
remains constant at all redshifts. To study this assumption further,
we consider cross-correlating with successively fainter magnitude
slices of the galaxy population. Since fainter galaxies, on average,
will probe higher redshifts it should in principle be possible to
constrain the redshift evolution of the galaxy population. Modifying
Equations 11 and 13 we can obtain theoretical cross-correlations of the
unresolved XRB with galaxies between the magnitude limits $[m1,m2]$:

\begin{equation}
\delta A_{xg}={ \rho_0 r_0^\gamma H_\gamma\over
 4\pi\bar{I}\Delta N }\int_z {\rm d} z (1+z)^{\aleph} r^{1-\gamma} \int_{m_1}^{m_2} 
N(m,z)\rm{d} m
\end{equation}

We can normalize this by the observed cross-correlation from the full
($18<B \leq 23$) galaxy sample, giving:

\begin{equation}
\delta A_{xg} =  {A_{xg}\bar{N} \over f(\aleph) \Delta N } 
\int_z {\rm d} z (1+z)^{\aleph} r^{1-\gamma} \int_{m_1}^{m_2} 
N(m,z)\rm{d} m
\end{equation}

Thus by performing the cross-correlation with galaxies in different magnitude
ranges  we may be able to constrain the evolutionary parameter $\aleph$ 
and hence the evolution in X-ray emissivity.

The galaxy catalogue was therefore split into ``bright'' ($18<B<
21$) and ``faint'' ($21 \leq B < 23$) subsets. The redshift
distributions for these samples are shown in Figure 5, as predicted
from the models of Roche et al. (1996a). Two separate
cross-correlations with the unresolved XRB are then carried out.  The
results are displayed in Figure 6. The brighter sample clearly show a
higher overall correlation amplitude. To compare with a range of
theoretical predictions (in order to constrain the evolution) we
evaluate the integral in Equation 17 to obtain expected 
amplitude $A_{xg}$ as a function of the evolutionary parameter
$\aleph$. These are displayed in Figure 7. Clearly we expect the
relative cross-correlation amplitudes to change dramatically depending
on the redshift evolution. With stronger evolution, the amplitude for
the brighter (more local) sample will decrease.

To constrain $\aleph$ we therefore evaluate $\delta A_{xg}$ for both
galaxy samples. We fit models of the form given by Equations 14 and
15.  A complication arises since the functional form we fit requires a
value for $q$ in advance (to evaluate the Poisson contribution), so we
start with the assumption that $q=0$ and proceed iteratively.  In
practice the $\em relative$ amplitudes of the Poisson and clustering
terms are fairly insensitive to the assumed value of $q$ and vary by
only $\sim 10\% $ with q in the range $q\in[0,3]$. Nevertheless we
start with a value $q=0$, use the measured amplitude $A_{xg}$ to
constrain the evolutionary parameter $\aleph$ (and hence $q$) and then
repeat the process.  Within a couple of iterations we converge on
values for the amplitudes $A_{xg}$ for both magnitude slices.

\begin{figure}
\centering
\centerline{\epsfxsize=8truecm 
\figinsert{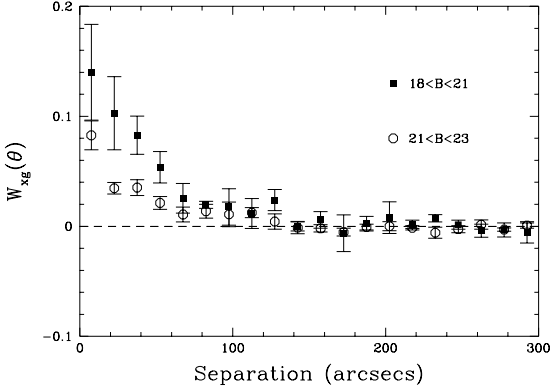}{0.0pt}}
\caption{The  cross-correlation function $W_{xg}(\theta)$
of galaxies with the unresolved $0.5-2.0$\,keV XRB from  three $\em ROSAT$ 
 fields. The galaxies are split into a bright  and faint sample.}
\end{figure}

\begin{figure}
\centering
\centerline{\epsfxsize=8truecm 
\figinsert{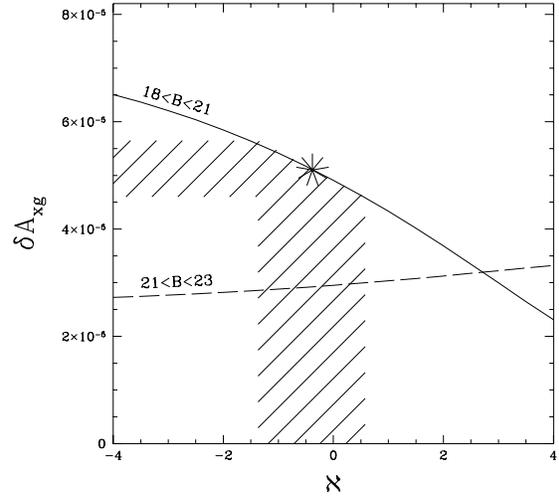}{0.0pt}}
\caption{Showing the expected relationship between the amplitude of the 
cross-correlation function $W_{xg}$ and the evolutionary parameter
$\aleph$ for brighter $18<B\leq21$ galaxies (solid upper line) and the
fainter $21 \leq B < 23$ galaxies (lower dashed line). The shaded
region displays the $1\sigma$ error bounds on the $\em measured$
amplitude for the brighter galaxies and the corresponding range of
$\aleph$. }
\end{figure}

With $\theta$ measured in radians we obtain amplitudes of $5.15 \pm
0.35 \times 10^{-5}$ for the bright galaxies and $2.8 \pm 0.2 \times
10^{-5}$ for the fainter sample. These amplitudes were obtained by
fitting to $W_{xg}(\theta)$ for $\theta < 2$ arcminutes and the errors
are derived from the variance in the amplitudes from 3 independent
fields. The resulting estimates for the evolutionary parameter
$\aleph$ are shown in Figure 7.  The shaded area shows the 1$\sigma$
confidence region on the $\em observed$ amplitude for the brighter
galaxies and the corresponding preferred values of $\aleph$.  For the
fainter sample however the weaker dependence of $A_{xg}$ on $\aleph$
does not allow us to provide any useful constraints on the
evolutionary parameter. This is expected however since, by
construction, Equation 17 is normalized by the amplitude of the full
$18<B <  23$ dataset and therefore dominated by these fainter
galaxies.

The errors on $\delta A_{xg}$ suggest values of $\aleph$ in the range
$-1.37 < \aleph < 0.58$.  This global parameter was defined $\aleph =
\gamma - \epsilon +q -5$ where $q$ describes the evolution of the
X-ray emissivity as a function of redshift (Equation 8) and $\gamma$
and $\epsilon$ characterise the clustering properties of the galaxy
population (see Equation 10). We have  assumed that the latter
two quantities are relatively well defined in comparison with the
parameter $q$ and adopt the values $\gamma=1.8$ and $\epsilon=0$.
This value of $\epsilon$ allows the galaxy evolution model of Roche et
al (1996a) to simultaneously fit the faint galaxy number counts and
the galaxy clustering.  This leads to values of $q=2.82 \pm 0.98$ and
thus a best estimate for the evolution of the X-ray emissivity due to
galaxies:

\begin{equation}
\rho_x(z) = \rho_o (1+z)^{2.82 \pm 0.98}
\end{equation}

Using a spectral index of $\alpha_x \sim 0.6$ (Almaini et al. 1996) we can obtain our best
estimate for the X-ray luminosity evolution of faint galaxies:

\begin{equation}
L_x \propto (1+z)^{3.22 \pm 0.98}
\end{equation}

which is very similar to the X-ray evolution of AGN.  Similar forms
have been obtained by considering the luminosity function of
individually identified X-ray luminous galaxies at much brighter
fluxes (Griffiths et al. 1995, Boyle et al. 1995a).

Since the $N(m,z)$ of Roche et al. (1996a) places the faint blue galaxy
population at relatively high redshift, it might be argued that the
strong evolution we have detected here is due to our particular choice
of galaxy evolution model. We have therefore repeated the analysis
using the model of Efstathiou (1995), following TL96, which assumes a
much more local population of faint galaxies. Although the mean
redshifts are lower, this model uses a much smaller correlation radius
($r_0=2~h^{-1}$Mpc) with comoving clustering evolution
($\epsilon=-1.2$) rather than the stable model we have used in this
paper ($\epsilon=0$). The end result is a very similar galaxy-galaxy
angular autocorrelation. Even with this model we still obtain evidence
for strong evolution in X-ray luminosity ($L_x \propto (1+z)^{ 3.5\pm
1.2}$).

\subsection{The X-ray volume emissivity of faint galaxies}

With an estimate for the X-ray evolution we may now estimate the
volume emissivity due to faint galaxies using Equation 11.  First we
require the amplitude of the observed angular cross-correlation
function $W_{xg}(\theta)$ (Figure 4).  As before, fitting the
functional form given by Equations 14 and 15, we obtain (measuring
$\theta$ in radians):

\begin{equation}
A_{xg}=3.60\pm 0.27 \times 10^{-5}
\end{equation}

The mean intensity of the unresolved XRB from the 3 deep $\em ROSAT$
fields used here is $\bar{I}= 1.59 \pm 0.2\times 10^{-8} $ erg
s$^{-1}$ cm $^{-2}$ sr$^{-1}$ and $\bar{N}=2.64 \times 10^7$ galaxies
sr$^{-1}$ in the magnitude range $18 < B < 23$.  Putting all of
this together we may now obtain the local X-ray volume emissivity via
Equation 11:

\begin{equation}
\rho_o\simeq 3.02\pm 0.23
\times10^{38} h\, {\rm erg s}^{-1}\rm{Mpc}^{-3}
\end{equation}

It should be emphasized however that this estimate relies on very
specific assumptions about the distribution and clustering properties
of the galaxy population.  In particular since $\rho_o \propto
r_o^{-\gamma}$ (from Equation 11), a less clustered model with
$r_o<4.4 h^{-1}$Mpc will $\bf increase$ the required local emissivity.
Such weak clustering would be required in models which place most
faint galaxies at relatively low redshift (eg. Efstathiou
1995). Recent deep spectroscopic work seems to confirm the existence
of a significant high redshift population however (eg. Cowie et
al. 1996).

\subsection{The contribution of faint galaxies to the unresolved XRB}

%
%


We may now obtain a rough estimate the total contribution of the
$18<B<23$ catalogue galaxies to the unresolved XRB by integrating the
volume emissivity out to the median redshift of the sample,
$z=0.45$. Modifying Equation 9 gives the following expression for the
contribution to the sky intensity per unit solid angle:

\begin{eqnarray}
\Delta \bar  I_g & = &  \omega \int_{z=0}^{z=0.45} {\rho_g (z) \over 4\pi r_l^2(z)}{\rm d}v(z) \\
\nonumber 
& = & 3.6 \pm 0.5 \times 10^{-9} {\rm erg s}^{-1} {\rm cm}^{-2} {\rm
sr}^{-2}
\end{eqnarray}

This accounts for 23$\pm 3\%$ of the unresolved X-ray background,
where the error is derived entirely from the uncertainty in the
observed X-ray quantities, $W_{xg}(\theta)$ and $\bar{I}$.

To determine the ``total'' galaxy contribution to the XRB we simply
extrapolate Equation 22 to the faintest galaxies at arbitrarily high
redshifts. This process becomes increasingly uncertain as we extend
the redshift distribution beyond the limit of our sample at $B=23$,
but nevertheless by integrating to $z=2$ we formally obtain $\sim
80\pm20 \%$ of the unresolved $0.5-2.0$\,keV XRB, which is $\sim 40
\pm10 \%$ of the total XRB intensity. Given the uncertainties already
inherent in this procedure it will probably suffice to say that
galaxies can produce a significant fraction of the XRB at least as
high as the contribution from QSOs.

\subsection{Red  and blue galaxies}

To constrain the colour and type of galaxies producing the
cross-correlation signal we use R band plates to separate the sample
into red and blue subsets, dividing at $B-R=1.5$. The
cross-correlations were then carried out separately for each
dataset. The results (Figure 8) show no significant difference in the
cross-correlation amplitudes, suggesting that galaxies of all
morphological types are contributing to the signal.  Blue galaxies at
these magnitudes are known to be more weakly clustered however, with
autocorrelation amplitudes lower by a factor of $\sim 4$ at these
magnitudes (Roche et al. 1996a) and hence the enhancement in
$W_{xg}(\theta)$ due to clustering will be significantly lower for the
blue subset. Thus the typical X-ray luminosities may be higher for the
blue galaxies (by $\sim 50\%$), but a more careful analysis is
required to prove this since the $N(m,z)$ distribution for blue and
red galaxies may be significantly different.

\begin{figure}
\centering
\centerline{\epsfxsize=8truecm 
\figinsert{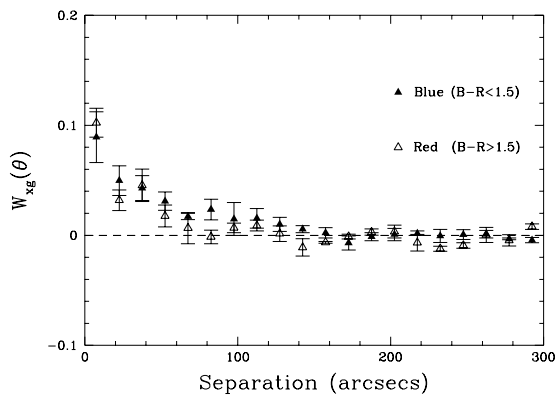}{0.0pt}}
\caption{The  cross-correlation function $W_{xg}(\theta)$
of galaxies with the unresolved $0.5-2.0$\,keV XRB, separating blue from 
red galaxies and combining the results  from  three $\em ROSAT$ 
 fields.}
\end{figure}

\section{Future prospects}

Having established the existence of a population of highly luminous
X-ray galaxies, the next intriguing question is the origin and nature
of this activity. One explanation could be the presence of a large,
hitherto undetected population of low luminosity Seyfert 1 galaxies.
This would certainly explain the cross-correlation signal.  So far
however the small number of individually identified X-ray luminous
galaxies have shown only narrow optical emission lines.  High
resolution optical spectroscopy suggests a mixture of starburst and
Seyfert 2 activity (Boyle et al 1995b, McHardy et al. 1997) but in
many cases the classification was ambiguous.  Iwasawa et al. (1997)
argue that the ambiguous line ratio diagnostics may be due to the
presence of an obscured active nucleus surrounded by significant
starforming activity in the host galaxy, with the AGN producing the
hard X-ray emission.  Spatially resolved optical spectroscopy would be
able to test this hypothesis. Obscured AGN provide a natural
explanation since the expected X-ray absorption can readily reproduce
the flat spectra of the XRB (Comastri et al. 1995, Madau et al. 1994).
At least two of the brighter X-ray galaxies identified in Almaini et
al. (1996) show clear evidence of X-ray obscuration. The discovery of
a high redshift counterpart to this population adds further credence
to this possibility (Almaini et al. 1995). Explaining the XRB with an
AGN population may be difficult however since QSOs are strongly
clustered and may violate the upper limits of the isotropy of the XRB
(Georgantopoulos et al. 1993, Soltan \& Hasinger 1994). Little is known
about the clustering characteristics of low luminosity and obscured
AGN however. 

Starburst activity alone is unlikely to explain the X-ray emission in
most of these galaxies. There have been suggestions that
massive X-ray binaries formed in the wake of star formation in early,
low metallicity epochs may provide a strong source of hard X-ray flux
(Griffiths and Padovani 1990, Treyer et al. 1992).  Such models have
difficulty in explaining the XRB however, since ASCA observations of
starburst galaxies reveal very  soft X-ray spectra which cannot
contribute significantly to the hard XRB (eg. Della-Ceca et al. 1996).
Many individually identified X-ray galaxies are also several
orders of magnitude brighter than any known starburst galaxy.  A third
explanation has emerged recently based on advection dominated
accretion onto quasars at low accretion efficiency (Di Matteo \&
Fabian 1997), but the enormous black hole masses required present
significant difficulties for this model.

We conclude that the obscured AGN hypothesis provides the most
plausible explanation for the origin of the X-ray background and the
bulk of the X-ray activity in these galaxies.  Infra-red spectroscopy
may provide a conclusive test of this model by allowing us to see
through the obscuring dust to detect broad emission lines in
moderately obscured AGN. If the nuclei are very heavily obscured then
the infra-red emission may also be obliterated, but in this scenario
it is difficult to produce the soft X-ray flux we observe with $\em
ROSAT$ without a very large $\sim 10 \%$ scattered component or some
additional source of soft X-ray photons. Spectrapolarimetry may then
allow us to detect broad emission lines scattered around the obscuring
medium and into our line of sight. Further observations are clearly
required before we can claim to understand the nature of this X-ray
activity.

\section{Summary and conclusions}

By cross-correlating faint galaxy catalogues with unidentified X-ray
sources a strong ($4.2 \sigma$) signal was detected indicating that
individual galaxies with magnitude $B<23$ account for $ 20 \pm 7 \%$
of all X-ray sources to a limiting flux of
$\sim4\times10^{-15}$erg$\,$s$^{-1}$cm$^{-2}$ in the $0.5-2.0$\,keV
band.  This builds on the results of Roche et al. (1995) who found a
significant signal in cross-correlation with brighter $B<21$ galaxies,
attributing $\sim 6\% $ of the X-ray sources to these brighter
objects.  Scaling the $20 \% $ galaxy fraction by the median flux of
the unidentified X-ray sources leads to $\sim 4 \% $ of the total
XRB intensity.

To probe deeper we cross-correlate with individual photons in the
remaining unresolved XRB images. Significant signals were obtained on
all 3 deep $\em ROSAT$ images, each of similar amplitude,
independently confirming the results obtained by Roche et al. (1995).
To translate these cross-correlations into a total fraction of the
unresolved XRB a specific description of the galaxy population is
adopted, modelling the evolution, number density and clustering
properties of the faint blue galaxy population using the formalism
developed by TL96.  We modify this formalism to allow for the effect
of finite cell sizes and the effect of the $\em ROSAT$ point-spread
function. By comparing the theoretical XRB cross-correlation with the
observed signal an estimate for the local X-ray volume emissivity was
obtained at $\rho_o\simeq 3.02\pm 0.23 \times10^{38} h\, {\rm erg
s}^{-1}\rm{Mpc}^{-3}$. Extrapolation to high redshift ($z=2$) suggests
that faint galaxies can account for $\sim 40 \pm 10 \%$ of the total
XRB at 1keV.  These estimates have a strong dependence on the assumed
distribution and clustering properties of the faint galaxy population.

When the optical galaxy catalogue is separated into blue and red
subsets, dividing at $B-R=1.5$, no significant difference is found in
the cross-correlation with the XRB. This would suggest that a mixture of galaxy
colours and morphologies contribute to the observed signal. Given that
red galaxies are more strongly clustered at these magnitudes, this may
also indicate that bluer galaxies are intrinsically more X-ray
luminous.

To constrain the evolution of X-ray emissivity with redshift, separate
cross-correlations were carried out with two magnitude slices of the
galaxy population. The resulting difference in amplitude suggests that
faint galaxies evolve strongly with redshift such that:

\begin{equation}
L_x \propto (1+z)^{3.22\pm 0.98}
\end{equation}

which represents the first evidence that the X-ray emission from faint
blue galaxies evolves as strongly as AGN. Similar results have been
obtained by analysing the brightest narrow emission-line galaxies
emerging from deep $\em ROSAT$ exposures (Griffiths et al. 1996, Boyle
et al. 1995a).

Combined with recent findings which suggest that X-ray luminous
galaxies have hard X-ray spectra, it now seems established beyond
doubt that a population of faint galaxies, or some processes
associated with them, are emitting vast amounts of X-ray radiation
which may finally solve the puzzling origin of the X-ray background.
Obscured AGN models provide a plausible explanation, but the true
nature of these X-ray luminous galaxies remains unclear.

\section*{ACKNOWLEDGMENTS}

OA was funded by a PPARC postdoctoral fellowship. OA thanks Marie
Treyer, Alexandre Refregier and Ofer Lahav for their considerable help
with this work. We also thank the referee, Guenther Hasinger, for helpful
and constructive comments.

\end{document}